\documentclass[12pt]{article}   
 \usepackage{psfig,times,fancyheadings}
        
\begin{document}
\thispagestyle{empty} 


 \lhead[\fancyplain{}{\sl }]{\fancyplain{}{\sl }}
 \rhead[\fancyplain{}{\sl }]{\fancyplain{}{\sl }}

 \renewcommand{\topfraction}{.99}      
 \renewcommand{\bottomfraction}{.99} 
 \renewcommand{\textfraction}{.0}


\newcommand{\nc}{\newcommand}

\nc{\qI}[1]{\section{{#1}}}
\nc{\qA}[1]{\subsection{{#1}}}
\nc{\qun}[1]{\subsubsection{{#1}}}
\nc{\qa}[1]{\paragraph{{#1}}}

\def\qbu{\hfill \par \hskip 6mm $ \bullet $ \hskip 2mm}
\def\qee#1{\hfill \par \hskip 6mm #1 \hskip 2 mm}

\nc{\qfoot}[1]{\footnote{{#1}}}
\def\qL{\hfill \break}
\def\qpar{\vskip 2mm plus 0.2mm minus 0.2mm}
\def\tvi{\vrule height 12pt depth 5pt width 0pt}

\def\qparr{ \vskip 1.0mm plus 0.2mm minus 0.2mm \hangindent=10mm
\hangafter=1}

\def\qdec#1{\par {\leftskip=2cm {#1} \par}}

\def\qdpt{\partial_t}
\def\qdpx{\partial_x}
\def\qddpt{\partial^{2}_{t^2}}
\def\qddpx{\partial^{2}_{x^2}}
\def\qn#1{\eqno \hbox{(#1)}}
\def\qds{\displaystyle}
\def\qw{\widetilde}
\def\qmax{\mathop{\rm Max}}   
\def\qmin{\mathop{\rm Min}}   


\null


\vskip 0.5cm
\vskip 1cm

\centerline{\bf \LARGE Comparing the correlation length}
\vskip 0.5cm
 \centerline{\bf \LARGE of grain markets in China and France}

\vskip 1cm
\centerline{\bf Bertrand M. Roehner $ ^1 $ }
\centerline{\bf L.P.T.H.E. \quad University Paris 7 }

\vskip 1cm
\centerline{\bf  Carol H. Shiue $ ^2 $ }
\centerline{\bf Department of Economics \quad University of Texas }

\vskip 2cm

{\bf Abstract}\quad  In economics comparative analysis plays 
the same role as
experimental research in physics.
In this paper we closely examine several methodological problems 
related to comparative
analysis by investigating the specific example of grain markets 
in China and France respectively. This enables us to answer a question 
in economic history which has so far remained pending, namely
whether or not market integration progressed in the 18th century. 
In economics as in physics, before being accepted any new result
has to be checked and re-checked 
by different researchers. This is what we call the replication and
comparison procedures. We show how 
these procedures should (and can) be implemented.

\vskip 1cm
 \centerline{\bf  8 November  2000 }

\vskip 1cm

Keywords: Correlation length; Comparative analysis; Grain markets;
          France; China.

\vskip 2cm
1: Postal address: LPTHE, University Paris 7, 2 place Jussieu, 
75005 Paris, France.
\qL
\phantom{1 }E-mail: ROEHNER@LPTHE.JUSSIEU.FR
\qL
\phantom{1 }FAX: 33 1 44 27 79 90

\vskip 1cm
2: Postal address: Dept. of Economics, University of Texas, 
Austin, TX 78712-1173, U.S.A.
\qL
\phantom{2 }E-mail: SHIUE@ECO.UTEXAS.EDU
\qL
\phantom{2 }FAX: 1 512 471 3510

\vfill \eject

\qI{Introduction: replication and comparison}

\qdec{Having gathered these facts, Watson, I smoked several pipes 
over them, trying to separate those which were crucial from others
which were merely incidental. \qL
Conan Doyle, The adventure of the crooked man (1975)}
\qpar

Experimental research has a dual role in physics 
(i) For new phenomena (for which there is no established
theory) it permits to separate factors which are of 
major importance from those which are not. In that kind of 
investigation comparative analysis plays a prominent role. 
(ii) For those
phenomena for which a theory exists it permits to check its 
predictions 
In both cases the cornerstone of the investigation is the replication
of crucial experiments under different conditions. For most
physicists (we exclude astrophysics from our discussion)
that double role of experiments would probably seem 
self-evident. 
\qL
In economics the analog of the operation of 
repeating an experiment consists in
carrying out consistent comparative observations. Unfortunately
economists have given very
little attention to the methodology of comparative analysis%
\qfoot{Some social sciences do not share that neglect: for instance,
in sociology the methodology of comparative analysis is a key issue
which has been extensively discussed (see for instance 
Mj\"oset 1997).}%
.
Even in the ``Journal of Comparative Economics'' there are few truly 
comparative studies. 
The lack of rigorous guidelines for comparative investigation has
detrimental consequences which are far-reaching as illustrated by the 
following examples. 

\qA{Physics versus economics: examples}
In physics, after a new phenomenon has been observed by a researcher
other groups around the world try to repeat the same experiment. If it 
cannot be repeated it will 
be attributed (possibly after a period of controversy) to some spurious or 
factitious effect and will be quickly forgotten. Such was for instance
the case of the ``discovery'' of cold fusion in 1989 by S. Pons and
M. Fleischmann. On the other hand, if the experiment can be repeated
and confirmed the phenomenon in question will become widely accepted
as a new building block of physics; fairly often this will open a new 
avenue of research. One can mention the following historic
examples (i) the Foucault pendulum (1851), (ii) 
the discovery of superconductivity by Kamerlingh Onnes
(1911), (iii) the discovery 
of the properties of spin glasses (mid-1970s). As an example of a
major discovery that has still to be confirmed, one can mention
the observation of a plasma of free quarks at the European
Center for Nuclear Research (announcement made in February 2000). 
A first confirmation (or refutation) will come within one or
two years from the experiments currently under way at 
Brookhaven National Laboratory (located on Long Island, New York State). 
A further check will be obtained in about five years by using the
Large Hadron Collider currently under construction at CERN. To sum up,
it may take some time, but within a few years the question of the
possible existence of a plasma of free quarks will probably be settled
in a satisfactory way.  

\qpar

In economics, in contrast,  controversies often
drag on for years without leading to any specific conclusion. 
A disenchanted but lucid
assessment of this state of affairs
has for instance been made in a review article by A. Zellner (1988). 
As an illustration one can mention the discussion around the 
so-called Prebisch - Singer conjecture (1950). Basically it posits 
a secular decrease in the price of primary commodities relative to the 
price of manufactured goods. In the last 50 years it has fostered 
numerous statistical tests but its validity still remains 
controversial (Hagen 1989, Cuddington et al. 1989).
\qpar

What are the reasons of such an unsatisfactory situation? Broadly speaking,
they gravitate around the two problems of {\it replication} 
and {\it comparison}. It is the purpose of this paper to examine these
questions more closely by carrying out a specific comparative study,
namely the comparison of geographically separated 
grain markets in China and Europe. For the 19th century, due to major
advances in transportation technology, it is fairly obvious that there
was a strong increase in market integration, but is that also true
for the 18th century? To our best knowledge that question has not
been clearly settled so far. In the present paper the approach that 
we use is almost as important as the answer that we arrive at. 
In the course of the investigation we will set forth a number of rules
that more generally can apply to the analysis of other comparative issues
as well. 

\qA{Replication in physics and in economics}
At first sight it could seem that the parallel with physics that 
we hinted to above is questionable. The obvious objection would be
that in physics one can repeat an experiment while in economics one
cannot. However a little reflection shows that such an objection does
not hold. As a matter of fact the central problems of 
replication and comparison present themselves very similarly
in both fields. Let us briefly see why. In physics these problems are
often condensed into the single question: ``Can a researcher $ B $
repeat the experiment carried out by a researcher $ A $?'' For 
definiteness let us consider the example of the Foucault pendulum. 
Replication would then mean: ``Knowing that Foucault has
carried out his experiment at the Panth\'eon in Paris, can I repeat
it at the same location and with the same pendulum?'' On the other
hand comparability would mean that the same observation can also
be made in Brazil with a different pendulum. By stating these questions
one immediately understands why they are usually considered together;
indeed nobody would try to perform {\it exactly} the same observation
as Foucault did. In the years after 1851 there were a number of experiments
with the Foucault pendulum in England, Germany and indeed in Brazil; 
not only were the locations different but the technical characteristics
of the pendulum (weight, length, nature and diameter of the wire) were
not the same either. 
By so doing one implicitly assumes that these 
characteristics are not crucial; that is possible
because the principles on which the phenomenon relies
are fairly well understood. 
As one
knows the outcome of the experiment is of course {\it }
not the same in Brazil
and in Paris but, knowing the theory, it is possible to 
take into account the effect of a different latitude%
\qfoot{Furthermore a spurious factor called the Puiseux effect had to 
be discounted. It states that if a pendulum describes an ellipse
(and all Foucault pendulums, whatever their technical
characteristics,
do in fact describe an ellipse) its major axis will turn at a rate
which is proportional to the area of the ellipse. It took some time
to experimenters to realize which correction had to be performed in 
order to discard that perturbation.}%
. 
\qpar

On the contrary in economics one does {\it not} know in advance which
factors are essential and which ones are of little importance. 
Suppose (for instance) one wants to 
examine if it is true that the distribution of income follows 
a Pareto law with an exponent of 1.4. For such a problem the
replication and comparison issues are clearly distinct one from
another. Replication means that if researcher $ A $ claims that 
result to be true for France in 2000, he (she)  must publish 
his (her) results
in such a way that researcher $ B $ can perform the same fit on the
same data and check if the same results obtain. This implies in 
particular that the data are made available to researcher $ B $, 
a condition which is only rarely fulfilled in practice. That 
obstacle to replication has been deplored by several lucid 
economists as for instance in the paper by Dewald et al. (1986). 
Comparability, on the other hand, would mean that if a researcher
performs a similar study for the distribution of
revenue in Germany (for instance)
he (she) will also find a Pareto exponent of 1.4. In general 
this will not be true of course, but in contrast to what 
happened with the Foucault pendulum
one does not know if the discrepancy should be attributed
to the measurement process (other kind of sources had to be used 
in Germany than in France)
or to the phenomenon itself. As a result
(in contrast to the 
experiment performed in Brazil) one is unable 
to apply to the German data whatever correction factor
which would be required in order 
to make them truly comparable to the 
French data. 
\qL 
To our best knowledge
\qfoot{May be some truly comparative investigations have
escaped our attention; we express in advance our gratitude
to those of our readers who would bring such studies to 
our attention. 
As a matter of fact we would be more than
pleased to see comparative analysis become a key issue in 
economic research.}
few
(if any) revenue investigations for different countries
are truly comparable; in  Atkinson's words (Smeeding et al. 1990)
``professor $ X $ will take the household as the unit of measurement,
while professor $ Y $ will take the nuclear family, and professor $ Z $
will take the revenue per head''. That would not be a problem if one
had at one's disposal a reliable correction formula in order to adjust
for family size; unfortunately this is not the case. 
\qL
As an illustrative example one can mention the 
``comparative'' study of income distribution edited by
Brenner et al. (1991): six countries are examined by six different
authors, but almost none of the data are comparable. The data for
Britain are based on wages (which excludes capital gain), those for 
Germany are based on data from tax authorities (which
include capital gains); 
in some cases a correction was performed which takes
into account the distribution of age groups while in others it was not,
and so on. 
\qL
Is the situation hopeless then? Certainly not. For one thing one
should consider a phenomenon which depends upon a few parameters only. 
That is why we selected the distribution of grain prices 
considered in this paper. Secondly, we will show that it is possible
to consider data which make  the comparison meaningful. 
\qpar

The paper proceeds as follows. In the second section we list the
parameters which affect the phenomenon under consideration; this will
help us to define data sets for which a meaningful comparison can
be made. In the third section we 
describe the
Chinese and French price data. In section 4 we show that the prices have
the same spatial structure in the two countries and we estimate the 
correlation length in each case. In section 5 we examine whether or not
market integration progressed in the 18th century. Finally in the 
concluding section we summarize
our findings along with the precepts which 
in a general way can improve the reliability of comparative analysis.

\qI{Comparing two systems of markets}
For a cross national comparison to be valid one must ensure that
all factors (except of course the factor
that one wants to study) are
identical. This is undoubtedly a very strong constraint; it is due to
the fact that
we do not know how to discount the effect of factors which are different.
A number of the factors that can be expected to be relevant for the
comparison of two systems of markets are listed in Table 1. 

\small

 \count101=1

 \ifnum\count101=1

\begin{table}[htb]

\centerline{\bf Table 1 \ Factors which affect the analysis of the spatial 
distribution of prices}
\vskip 3mm
\hrule
\vskip 0.5mm
\hrule
\vskip 2mm
$$ \matrix{
\tvi \hbox{\bf Well-defined parameters} \hfill &  &  \cr
 & & \cr
 & 1 & \hbox{Unit of currency} \hfill \cr
 & 2 & \hbox{Unit of weight or volume} \hfill \cr
 & 3 & \hbox{Type and quality of grains} \hfill \cr
 & 4 & \hbox{Type of prices (market/contract prices)} \hfill \cr
 & 5 & \hbox{Frequency of prices} \hfill \cr
 & 6 & \hbox{Size of the region} \hfill \cr
 & 7 & \hbox{Topography of the region (plain/mountain/sea-coast)} \hfill \cr
 & & \cr
\hbox{\bf Broad factors} \hfill &  &  \cr
 & & \cr
 & A & \hbox{Propensity for trade} \hfill \cr
 & B & \hbox{Means of transportation} \hfill \cr
 & C & \hbox{Business situation} \hfill \cr
} $$

\hrule
\vskip 0.5mm
\hrule

\vskip 2mm
Notes: For a comparative study to be meaningful 
all factors must be identical (or at least controlled in the
sense that a reliable correction can be performed) except the
one under study. That is why this table will subsequently be referred
to as the ``ceteris paribus'' (i.e. everything else being identical)
table. 

\end{table}

\fi

\normalsize

Needless to say, this list cannot be exhaustive; one has to posit that
all factors that are not explicitly considered play a negligible
role. It is only by performing numerous comparative studies for the same
phenomenon that one can separate the factors which are important 
from those which play only a small role. In fact, the
situation is the same in physics: when an experiment is carried out for the
first time in a new field one does not know which factors are 
crucial and which ones are not. 
\qpar

In table 1 we made a distinction between factors (labeled 
by numbers) which are fairly well defined and can therefore be
easily controlled, and more complex factors (labeled by capitals).
A few words are required to explain the phenomena to which these
entries refer. What do we mean by ``propensity for trade''?
Propensity for trade will for instance be low for fragile goods such
as eggs or fresh figs, or for goods which 
cannot be stored for a long time, or
in a general way for all products for which transportation costs 
are too large compared to their value. Thus, one would expect
the propensity for trade to be larger for caviar than for wheat. 
Another (less exotic) example is wheat versus potatoes. In 19th 
century Germany one kilogram of wheat was worth at 
least three times as much
as one kilogram of potatoes%
\qfoot{The calculation which leads to that estimate goes as 
follows. In the decade 1891-1900 the average price of wheat in 
Berlin was 164 mark/ton (Jacobs et al. 1935), while the average
price of potatoes was 2.19 mark/50 kg that is to say 43.8 mark/ton,
i.e. 3.7 times less than the price of wheat. Needless to say since
the prices of wheat and potatoes fluctuate in a fairly uncorrelated
manner this price ratio fluctuates as well: 
it was equal to 4.4 in the decade 1811-1820, and to 5.3 in 1925-1934.
For the United States, in the second half of the 20th century
that ratio seems to be much lower, of the order
of 1.2 to 2.}%
.
\qL
Regarding entry B, a central question is whether a decrease in
transport costs will lead to an increase in trade
and hence to a stronger market-interdependence. Although 
the answer could at first 
sight seem fairly obvious, it is not in fact. Indeed, a 
decrease in transport cost between two cities $ A $ and $ B $ leads
to a decrease in the price differential $ p_A -p_B $, and in the
face of such a lower price differential traders will have a lower
incentive to trade (assuming that their profit rate is somehow 
determined by the price differential). To decide which of these
opposing effects will prevail is not obvious. In the framework of the
stochastic spatial arbitrage model 
(Roehner 1995,1996) it is found that a decrease
in transport cost leads to a trade increase,
a prediction which seems to be confirmed by empirical evidence
(Roehner 1995, p.151, and 2000, p.193)%
\qfoot{Once again it is 
because of the ceteris paribus requirement that 
the question is not easy to settle empirically.}%
. 
\qL
The business situation (that is to say whether or not one is in
a phase of recession, whether interest rates are low or high,
and so on) 
undoubtedly plays a role, but in a way which is difficult
to specify. Usually, one expects that by 
considering a sufficiently large span of time (e.g. 50 years or more)
these effects can be averaged out. 
\qL
Note that the previous factors could be replaced by a single variable which
is the trader's profit margin. Unfortunately 
that kind of information is rarely (if ever) made public and cannot 
be controlled therefore; this is why one must contend oneself with the
indirect criteria listed above. 
\qpar

The previous discussion could well lead to a pessimistic view. 
In the previous section
we claimed that the present problem has been
selected on account of its relative simplicity, and yet the above 
discussion seems to show that it has in fact numerous facets. 
To get a more realistic perception let us once again consider the
example of the Foucault pendulum. Even if the experiment is 
repeated on two successive days, at the same place and with the
same device one could well {\it argue} that the conditions 
have changed. Indeed the position of the Moon (which undoubtedly
exercises a gravitational attraction on the pendulum) has changed
as well as (though to a smaller extent) the position of the main
planets. Furthermore, vibrations transmitted to the pendulum
from the outside world are not the same on two successive days.
The physicist would answer that the gravitational attraction of the
Moon is negligible (it is about $ 10^6 $ times
smaller than the attraction of the earth)
while outside vibrations will be averaged out. Well and good,
but the fact that the precision of the measurement does not exceed
5 percent in the best experiments and is more often of the order
of 10 percent clearly shows that there are indeed a number of 
perturbations which are not well understood and hence cannot be
controlled. In spite of such defects the experiment is nevertheless
highly successful in demonstrating the rotation of the earth. 
Similarly, for the economic problem under consideration, one  can
hope that in spite of the fact that there are some poorly
controlled factors, one will nevertheless get a reliable 
comparison. 

\qI{Grain prices}
In contrast to many aggregated macro-variables commodity prices 
are well defined and can be accurately measured. This does of course
not mean that comparisons of price data are ipso facto reliable. As 
an illustration one can mention the monthly share prices published
by the Organization for Economic Cooperation and Development (OECD 1966,
1989): prices for the Milan stock exchange are averages
of daily quotations, figures for Oslo give the quotation on
the 15th of each month, the data for Stockholm are quotations at the
end of each month. Clearly such figures are {\it not} comparable%
\qfoot{Assuming independent random price fluctuations,
the standard deviation of the Milan prices will be five times
(i.e. $ \sqrt{30} $) times smaller than that of the two other markets.}%
. 
Such a lack of comparability in an official publication is all the more
surprising when one considers that daily share price statistics are easily 
available; if anything, it proves that data-comparability is not of
major concern for international statistical organizations. 
\qL
For 18th century grain prices there are usually two main sources
(i) Prices recorded by government officers on each grain market of
some importance (ii) Prices at which grains are bought by various
institutions such as hospitals, monasteries, and so on. Prices from the
second source usually display a smaller volatility (i.e. standard 
deviation) than prices from the first, but this difference
is sizeable only when one considers monthly or weekly prices. 
For the annual prices that we use in this paper the two kinds of
prices would be fairly comparable. However, since we need a comprehensive
geographical coverage the prices recorded by government officers
will be more homogeneous. The prices that we use subsequently are of that
kind; let us describe them in more detail. 

\qA{China}
At least from the 17th century, a systematic reports of the prices of
the major grains was required from every county (\textit{xian}) magistrate.
Prices were recorded at a minimum interval of once per month. The local
government of the county was charged with the task of investigating the
markets within or serving the main city of the county and recording the
selling price of different grains. These reports were subsequently sent
to
the prefectural city (\textit{fu}). The county reports were then
summarized
by two prices for each grain, the highest and lowest price among all 
the
counties. The county reports and the summary were then sent on to the
provincial capital, where the provincial governor then used the summaries
to
prepare monthly price reports to the Emperor. While the county level
reports
have been largely destroyed, copies of the monthly price report summaries
at
the prefectural level have survived, and it is these price reports, the
Grain Price Lists in the Agricultural Section of the Vermillion Rescripts
in
the Palace Archives [\textit{Gongzhong shupi souze, nonye lei, liangjia
qingdan}], currently kept in the Number One National Archives of
Beijing,
which are used in this paper.
\qL
There were two main reasons why the government collected price data
along
with local weather reports and local harvest reports. First, to deal
with
problems of mass riots resulting from food shortages, the government
maintained a sophisticated system grain management. Disaster relief
institutions included tax relief or tax postponement, cash and grain
transfers to local governments, seed loans, and grain disbursal from
public
granaries. Price data was used to help monitor the market conditions,
the
local grain supply availability, and the harvest outcomes in local
areas.
The information was used to preempt potential crises resulting from
grain
shortage, and, to assess the severity of food shortage and thus the
validity
of applications for food relief from local officials. Second, the
government
had to purchase grain for the consumption of its soldiers and the
approximately one million bureaucrats residing in Beijing. Price 
reports
allowed government officials to identify which market to purchase from. 
The
fact that the government utilized the price data for comparative
purposes
strongly suggests that not only is the accuracy of the prices 
relatively
high, but that the prices have been converted into units of silver
{\it tael }which are comparable in value%
\qfoot{Prices of grain would have been posted 
locally in terms of a copper cash
price, whereas the standard government unit of price accounting
converted
all copper prices to units of silver {\it tael }(Wilkinson,
1969)}%
.  
The reports could not have served their purposes if they were not
in comparable units of currency across regions.

\qA{France}
In France the recording of grain prices goes back to a royal 
ordinance of 1539. One may wonder why the government monitored so
closely grain prices. (i) Grains constituted the main component
in the diet of the population; any sudden price increase
could lead to riots. (ii) When troops were marched through the 
country the government had to know at which price grains were
available. In short, grain prices were as important in that age
as oil or stock prices are nowadays. 
\qL
The 18th century price series that we use were built for the
information of the National Convention (1792). They were 
subsequently re-published in 1837 (Archives Statistiques) and 1933
(Labrousse). The series give average annual prices for each 
of the 32 regions composing France at that time. We do not really
know how the averages were computed, that is to say which markets
were used in each region to define the region-average; however,
some tests performed by Labrousse (1933, p.70) show that the 
data are consistent with control-averages performed using 4 or 5 markets
(which means that with 5 terms the convergence towards the expectation
is already good enough). 
\qL
An important point concerns units of measurement. As one knows,
back in the 18th century there was a great diversity as to 
units of measurement. Yet, in the source of 1792 all prices
are expressed in livres per ``setier de Paris''. How were the
conversions performed? In 1755-1756 a systematic survey was 
carried out from which the required conversion factors could be
drawn (Labrousse, 1933). 

\qI{Measuring the correlation length}
As in statistical mechanics the correlation length, $ L $, of a 
set of markets characterizes the range of the interaction. 
It is defined through the formula: $ \rho (d)=e^{-d/L} $
which expresses the correlation $ \rho $, between prices on two markets
as a function of the inter-market distance $ d $.
The operational
definition of the correlation length is recalled in Fig.1. 
Note that for the sake of convenience we will rather use
a smaller correlation length $ l $ defined by: $ l=L/100 $
It should be emphasized that the correlation length is in many
respects ideally suited to the structural comparison of two 
systems of markets. Because the correlation between two price 
series is independent of the respective magnitude of the prices,
conditions 1,2,3 in Table 1 are automatically satisfied; and 
since the correlation length characterizes the behavior of the
correlation as a function of distance the overall size of the
region under consideration (condition 6) does not matter. 
\qL
In order to ensure that one compares two territories characterized 
by a similar topography (condition 7) one must restrict the analysis
to regions of limited extent. As a result a global comparison
of China and France would be questionable; indeed, as each country has
plains as well as mountains, the result would be an ill-defined average
whose interpretation would be open to debate.
\qL
A last remark is in order. In the same way as one expects price 
correlations to decrease with intermarket distance one would expect price
differentials to increase with distance. 
Why then did we not also study 
price differentials; in fact we did, but 
that dependence turns out to be very noisy and chaotic. 
An obvious reason 
for the smoother behavior of correlations 
is the fact that the latter implicitly involve
a built-in time average. If adequate time-averaging is performed 
the differential - distance relationship becomes meaningful
as well (see Roehner 1995, p.133-137).

\qA{China (Jiangsu)}
The most developed economic region of China is generally regarded to be
in
the Yangzi Delta, a plains area on the eastern coast in the provinces 
of
Jiangsu and Zhejiang and situated at the center of the most important
long-distance water routes of the 18th century: the mouth of the Yangzi
River, the Grand Canal, and a sea port.  The province of Jiangsu is
significant because all three of the major water routes, and many
tributaries, pass through the province, giving it a natural advantage 
in
transport costs. 
Table 2a shows the correlation length for two samples. Sample 1 consists
of
all ten prefectures of Jiangsu province for the series 1742-1795.
Inter-market distances are measured from the capital cities of each of
the
prefectures. For every 4.8 kilometer increase in distance between markets
in
Jiangsu, the correlation declines by one percent. Figure 3a graphs the
relationship between the log of correlations and distance in kilometers. 

\small

\begin{table}[htb]

\centerline{\bf Table 2a \ Correlation length for Chinese grain
markets 1742-1795}
\vskip 3mm
\hrule
\vskip 0.5mm
\hrule
\vskip 2mm
$$ \matrix{
\tvi & \hbox{Sample} & \hbox{Regions (prefectures)} \hfill & 
\hbox{Correlation length} \cr
 & \hbox{} & \hbox{} & \hbox{[km]} \cr
 & \hbox{} & \hbox{} & \hbox{} \cr
\noalign{\hrule}
\tvi
 1 & \hbox{Jiangsu} & \hbox{Changzhou, Haizhou, Huaian, Jiangning, Sungjiang } 
\hfill  & 4.8 \pm 2.3 \cr
   & \hbox{} & \hbox{Suzhou, Taicang, Tongzhou, Yangzhou, Zhenjiang}
\hfill & \cr  
 & \hbox{} & \hbox{} &  \cr
 2 & \hbox{Yangzi Delta} & \hbox{Changzhou, Hangzhou, Huzhou, Jianxing}
\hfill & 4.3 \pm  5.9 \cr
   & \hbox{} & \hbox{Ningbo, Shaoxing, Sungjiang, Suzhou}
\hfill & \cr 
} $$

\hrule
\vskip 0.5mm
\hrule

\vskip 2mm
Notes: In each region the geographical center has
been identified with the capital city. 
The results in this table refer to rice prices. It turns out that
wheat and rice prices 
(data for both cereals are available for
Hunan)
per unit of volume             
or per unit of weight          
(one liter of rice weights approximately 81 kg, Chuan et al. 1975)
were fairly comparable which means that wheat and
rice have almost the same propensity for trade. 
The error bounds were not obtained 
in the same way as in the case of France and may
be somewhat over-estimated. 

\end{table}

\normalsize

\qA{France}
Fig.3b shows the decrease of the correlation as a function of distance
for a sample of 10 regions. Subsequently this sample will be referred to
as the long-term sample because the data for these regions can be 
extended to the 19th century. Note that the very fact that there is a
well-defined relationship between correlations and inter-market distances
is a non-trivial property of the system of markets. It is this 
structural property which allows
a correlation length to be defined. 
For the long-term sample the correlation length is $ l=16.5 \pm 3.3 $ km. 
One
must of course examine to what extent that result is modified when one
considers another sample. Table 2b provides the result for three different
samples. 

\small

\begin{table}[htb]

\centerline{\bf Table 2b \ Correlation length for French grain
markets 1756-1790}
\vskip 3mm
\hrule
\vskip 0.5mm
\hrule
\vskip 2mm
$$ \matrix{
\tvi & \hbox{Sample} & \hbox{Regions} \hfill & \hbox{Correlation length} \cr
 & \hbox{} & \hbox{} & \hbox{[km]} \cr
 & \hbox{} & \hbox{} & \hbox{} \cr
\noalign{\hrule}
\tvi
 1 & \hbox{Long-term} & \hbox{Alen\c con, Amiens, Bourges, Bourgogne,} 
\hfill  & 16.5 \pm 3.3 \cr
   & \hbox{sample} & \hbox{Bretagne, Caen, Lyon, Riom, Rouen, Tours}
\hfill & \cr  
 & \hbox{} & \hbox{} &  \cr
 2 & \hbox{Sample 2} & \hbox{Alsace, Bordeaux, Champagne, Franche-Comt\'e,}
\hfill & 16.1 \pm 4.1 \cr
   & \hbox{} & \hbox{La Rochelle, Limoges, Metz, Poitiers, Soissons}
\hfill & \cr 
 & \hbox{} & \hbox{} &  \cr
 3 & \hbox{Sample 3} & \hbox{Auch, Bordeaux, Champagne, Franche-Comt\'e,}
\hfill & 12.8 \pm 2.3 \cr
 & \hbox{} & \hbox{Languedoc, La Rochelle, Limoges, Montauban, Poitiers}
\hfill &  \cr
} $$

\hrule
\vskip 0.5mm
\hrule

\vskip 2mm
Notes: For the sake of homogeneity we have used French spellings 
even when the English spelling was different (as for instance for Lyons)
The first two samples are restricted to the
northern part of France, while the third also includes locations
from the south (e.g. Auch, Languedoc, Montauban). The fact that the
third estimate of the correlation length is somewhat different is hardly
surprising since it does not correspond to the same region; it gives
a measure of the variability of the correlation length as one shifts
from northern to southern France. The geographical center of each
region has been defined as the localization of the capital region; for 
instance for Alsace it is Strasbourg, for Franche-Comt\'e it is
Besan\c con, and so on. The error bars correspond to the confidence
intervals at probability level 0.95. 

\end{table}

\normalsize

Sample 2 which is made up of non-mountainous regions 
located in the northern part of France leads to a correlation 
length which is almost identical to the previous one.
Sample 3 consists also of 
non-mountainous regions but they are located in the northern as well as
southern part of France; in this case the correlation length is
somewhat lower which suggests that southern France had 
a lower degree of market integration than northern France. 
\qL
The correlation length estimates obtained in this section provide
a static overall picture of market integration for the second half of the
18th century, but what about the evolution of market integration?
To analyze changes in the  
correlation length between 1750 and 1790 one 
would have to use a narrower time-window; yet,
in order to estimate the correlation with acceptable precision 
the time-window must contain at least 40 - 50 prices. Thus, to get a 
dynamical picture one would need monthly or weekly prices. 
As no high frequency data are available one must use a different
approach. This is discussed in the next section.

\qI{Did market integration progress in the 18th century?}
In  order to estimate the degree of market integration for a system
of $ n $ markets in a given year $ t $ we propose three different
measures.
\qbu The difference between the logarithm of the maximum and minimum
prices within the set of $ n $ markets; if the prices on the $ n $ 
markets are denoted by $ p_i(t) $ this variable reads
$$ a(t) = \qmax _{1 \leq i \leq n}[\ln (p_i(t)] -
          \qmin _{1 \leq i \leq n}[\ln (p_i(t)] $$

In statistics this difference (without the
logarithms) is called the range of the sample.

\qbu The (spatial) standard deviation of the logarithms of the prices
on the $ n $ markets:
$$ b(t) = \sigma [ \ln p_1(t),\ln p_2(t), \ldots , \ln p_n(t) ] $$

\qbu The (spatial) coefficient of variation:
$$ c(t)= { \sigma [ p_1(t), p_2(t), \ldots , p_n(t) ] \over
        m [ p_1(t), p_2(t), \ldots , p_n(t) ] } $$

where $ m [ p_1(t), p_2(t), \ldots , p_n(t) ] $ denotes the average price
on the $ n $ markets. 
\qL

Let us briefly explain what motivated the choice of these variables.
Firstly, it must be noted that $ a, b, c $ are scale invariant in the
sense that they do not change when all prices are multiplied by 
the same constant. As a result, these variables are independent of the
choice of the units of volume and currency which makes them well suited
for cross-national comparisons. Note that whereas the correlation was
``naturally'' scale invariant, one has here to use log-prices in order
to achieve that invariance (at least for $ a $ and $ b $). Regarding
the definition of $ a $ its main advantage is the fact that 
it can be computed very easily (almost by mere inspection of 
the price series); on the other hand it does not use the price
data very efficiently since it takes into account only two prices. 
Moreover, the confidence interval of the range $ a $ is known to
be fairly large: for large samples it decreases as $ 1 / \sqrt{\ln n} $ 
instead of the usual $ 1/\sqrt{n} $ (Kendall et al. 1987, p.461). 
\qL
In what follows we want to study the evolution of the above variables
in the course of time, in particular we examine whether or not
there is a downward trend. However, it must be noted that if there
is a downward trend it cannot be linear because $ a,b,c $ are 
necessarily positive. As a result a linear regression will not
give a good fit. A remedy is to fit instead: $ A=\ln a,\ B=\ln b,\
C=\ln c $; since $ A, B, C $ are not bounded they can have
have a linear downward trend in the course of time.
More specifically, the regressions performed in the 
following paragraphs correspond to the determination of the 
coefficient $ \alpha $ defined as (similar expressions for $ B, C $): 
$$ A=-\alpha t + \beta \Longrightarrow a=Be^{-\alpha t} $$

by their form these regressions parallel in the time domain
the regressions previously performed in the spatial
domain. 

\qA{China (Jiangsu)}

Table 3a gives the results for the regressions with respect to time.
For the Jiangsu province the regression coefficients are not 
significantly different from zero (except perhaps the third);
this is confirmed by the fact that the different criteria lead to
conflicting signs. On the other hand for the Yangzi Delta the
regression coefficients are significantly different from zero; 
this is confirmed by the fact that
the three criteria lead to consistent signs.  

\small

\begin{table}[htb]

\centerline{\bf Table 3a \ Trend $ \alpha $ for market integration in
China: 1742-1795}
\vskip 3mm
\hrule
\vskip 0.5mm
\hrule
\vskip 2mm
$$ \matrix{
\tvi & \hbox{Sample \hfill} & \hbox{Regression for}\ A 
& \hbox{Regression for}\ B & \hbox{Regression for}\ C \cr
& \hbox{} & [\hbox{century}^{-1}]
& [\hbox{century}^{-1}] & [\hbox{century}^{-1}] \cr
\noalign{\hrule}
\tvi
 1 & \hbox{Jiangsu \hfill} & 0.04 \pm 0.1 & 0.012 \pm 0.02
& -0.16 \pm 0.08 & \cr
 2 & \hbox{Yangzi Delta \hfill} & -0.18 \pm 0.1 & -0.04 \pm 0.02
& -0.22 \pm 0.06 & \cr
} $$

\hrule
\vskip 0.5mm
\hrule

\vskip 2mm
Notes: The samples are the same as in table 2a. $ A, B, C $ 
respectively denote the Max-Min, standard deviation of log-prices
and the coefficient of variation (see text). 

\end{table}

\normalsize

\qpar
\def\ql{\hbox{,lo}}
\def\qh{\hbox{,hi}}
\def\qwp{\hbox{wp}}

Table 3a does not make full use of all data that are avaible for 
China. Indeed (see section 3.1) for each prefecture $ (i) $ the
archives give the lowest ($ p_{i\ql} $) and highest ( $ p_{i\qh} $)
price among all the counties composing prefecture $ i $. 
So far, for each prefecture, we used the average price 
$ p_{i,\hbox{m}}=(p_{i\ql}+p_{i\qh})/2 $ (that we simply denoted $ p_i $).
However the $ p_{i\ql} $ and $ p_{i\qh} $ prices can give us
some information about the distribution of prices at a within-
(or infra-) prefecture scale. To this end we tentatively introduce
the following variables. 
\qL
(i) The difference between the logarithm
of the highest and lowest price within prefecture $ (i) $ averaged
over all  $ n $ prefectures in a given region:
$$ \alpha _{\qwp}={ 1 \over n}\sum _{i=1}^{n}(\ln p_{i\qh} - \ln p_{i\ql}) $$

(ii) The standard deviation of the logarithms of the lowest/highest
price averaged
over all  $ n $ prefectures in a given region:
$$ \beta _{\qwp}={ 1 \over n}\sum _{i=1}^{n} \sigma (\ln p_{i\ql},
\ln p_{i\qh}) $$

(iii) The ratio of the standard deviation of the lowest/highest price 
to the average price 
averaged
over all  $ n $ prefectures in a given region:
$$ \gamma _{\qwp}={ 1 \over n}\sum _{i=1}^{n} \sigma (\ln p_{i\ql},
\ln p_{i\qh})/p_{i,\hbox{m}} $$

{\bf Remark} The analog of the variable $ a(t) $ for this data set would
have been:
$$ a'(t) = \qmax _{1 \leq i \leq n}[\ln (p_{i\qh} (t)] -
          \qmin _{1 \leq i \leq n}[\ln (p_{i\ql} (t)] $$

but since this data set does not permit to compute the analogs of $ b(t),
c(t) $ we rather adopted the above definitions for the sake of homogeneity.
\qL
The regressions for $ \alpha _{\qwp}, \beta _{\qwp}, \gamma _{\qwp} $
are given in table 3a'; all coefficients are negative and significantly
different from zero at 5 percent significance level which
non-ambiguously attests to a 
downward trend. In other words progress of market
integration seems more pronounced at smaller distances (of the order '
of 50 km) than at distances of several hundredths kilometers. 

\small

\begin{table}[htb]

\centerline{\bf Table 3a' \ Trend $ \alpha $ for market integration in
Jiangsu within prefectures: 1742-1795}
\vskip 3mm
\hrule
\vskip 0.5mm
\hrule
\vskip 2mm
$$ \matrix{
\tvi & \hbox{Sample \hfill} & \hbox{Regression (i)} 
& \hbox{Regression for (ii)} & \hbox{Regression for (iii)} \cr
& \hbox{} & [\hbox{century}^{-1}]
& [\hbox{century}^{-1}] & [\hbox{century}^{-1}] \cr
\noalign{\hrule}
\tvi
  & \hbox{Jiangsu \hfill} & -0.0071 \pm 0.002 & -0.006 \pm 0.003
& -0.026 \pm 0.008 & \cr
} $$

\hrule
\vskip 0.5mm
\hrule

\vskip 2mm

\end{table}

\normalsize

\qA{France}
Table 3b gives the results for the regressions with respect to time.
\small
%
\begin{table}[htb]

\centerline{\bf Table 3b \ Trend $ \alpha $ for market integration in
France: 1756-1790}
\vskip 3mm
\hrule
\vskip 0.5mm
\hrule
\vskip 2mm
$$ \matrix{
\tvi & \hbox{Sample \hfill} & \hbox{Regression for}\ A 
& \hbox{Regression for}\ B & \hbox{Regression for}\ C \cr
& \hbox{} & [\hbox{century}^{-1}]
& [\hbox{century}^{-1}] & [\hbox{century}^{-1}] \cr
\noalign{\hrule}
\tvi
 1 & \hbox{Long-term sample \hfill} & -0.72\pm 0.95 & -0.87 \pm 0.96 
& -0.89 \pm 0.94 & \cr
 2 & \hbox{Sample 2 \hfill} & -0.39 \pm 1.10 &  -0.39 \pm 1.10 
& -0.31 \pm 1.11 & \cr
 3 & \hbox{Sample 3 \hfill} & -1.25 \pm 1.11 &  -0.94 \pm 1.05 
& -1.11 \pm 1.01 & \cr
} $$

\hrule
\vskip 0.5mm
\hrule

\vskip 2mm
Notes: The samples are the same as in table 2b. $ A, B, C $ 
respectively denote the Max-Min, standard deviation of log-prices
and the coefficient of variation (see text). The figures give the 
slope of the regression lines with respect to time
expressed in centuries. As shown by 
the width of the confidence interval the three variables fluctuate
markedly in the course of time. 

\end{table}
\normalsize
First it can be noted that the variables $ A, B , C $ lead 
to close results. Secondly, the width of the confidence interval 
shows that they are highly fluctuating in the course of time 
(this can also be seen on Fig.4b). The fact that all samples 
lead to negative regression coefficients shows that the 
downward trend is a robust property. This leads us to the conclusion
that both in northern and southern France market integration
made notable progress in the forty years preceding the Revolution
of 1789. The rate of the downward trend is between 3 to 8 times
faster (according to which sample one considers) than the one
observed in the Yangzi Delta region.
\qL
Naturally, because of the transportation
revolution, one would expect that
market integration has progressed even faster in the 19th century. 
This is indeed confirmed by Fig.4b. The decrease rate is almost twice
as large in the 19th century than it was in the second half of the
18th century. 

\qI{Conclusions}

Two main results emerge from this comparative study  
(i) In
the 18th century the level of market integration in France (expressed by
the correlation length of grain prices) was about three 
times as large
as in the plain region of Lower Yangzi, on the eastern coast of China.
(ii) Between 1750 and 1800 there was  substantial progress of market
integration in France; there was parallel progress in the Yangzi Delta
and even in Jiangsu province if one takes into 
account the sub-prefecture distance scale.
\qL
Needless to say, it would be of great interest to complement
these results with data pertaining to northern China and in particular
to the prefectures around Beijing.
\qL
The study also lead to other findings, such as for instance
the fact that 18th century grain 
markets were less integrated in 
southern than in northern France, or the fact that for France
progress of market integration in the
19th century was about twice as fast 
as in the 18th century. 
\qpar

But, beyond these specific results, the main message of this paper
was to advocate the definition and
adoption of rigorous guidelines for comparative
analysis. 
In particular, we emphasized that (i) at an experimental level 
the situation is not fundamentally different in physics and in 
economics (ii) very little attention has been devoted by 
economists to the definition of metholological rules for comparative 
analysis.
Hopefully, the present study
can serve as a starting point for other cross-national comparisons.

\vfill \eject

 \appendix

\qI{Appendix A: Price data}

It is of crucial importance that other researchers can check
our results and convince themselves that our conclusions do not
depend on built-in artifacts
at the level of data selection
or statistical analysis. This is what we called the replication
requirement. For that purpose the present appendix provides the
primary price data we have used. 
The Chinese data have never been published and are available only
from the archives in Beijing; the French data have been published
in a French thesis (Labrousse 1933) which may be difficult to obtain.
In order to save space we 
restricted the Chinese data to Jiangsu and 
the French data to the 18th century section of the long-term sample.
The data for the other samples 
are available from the authors on request. 

 \scriptsize

\begin{table}[htb]

\centerline{\bf Table A1 \ Rice prices in Jiangsu 
(Central-East China): 1756-1785}
\vskip 3mm
\hrule
\vskip 0.5mm
\hrule
\vskip 2mm
$$ \matrix{
\tvi \hbox{Date} & \hbox{Jiangning}& \hbox{Suzhou} & \hbox{Sungjiang} &  \hbox{Changzhou} & \hbox{Zhenjiang} & \hbox{Huaian} &
 \hbox{Yangzhou} & \hbox{Taicang} & \hbox{Haizhou} & \hbox{Tongzhou} \cr 
1742.2 &1.40 &1.43 &1.31 &1.38 &1.44 &1.47 &1.50 &1.54 &1.62 &1.49 \cr
1742.7 &1.49 &1.76 &1.73 &1.70 &1.60 &2.03 &1.70 &1.95 &1.73 &1.77 \cr
1743.2 &1.45 &1.53 &1.47 &1.55 &1.53 &1.67 &1.59 &1.62 &0.88 &1.61 \cr
1743.7 &1.40 &1.49 &1.47 &1.40 &1.46 &1.35 &1.25 &1.60 &2.05 &1.46 \cr
1744.2 &1.35 &1.62 &1.53 &1.47 &1.47 &1.38 &1.30 &1.75 &2.00 &1.43 \cr
1744.7 &1.30 &1.38 &1.52 &1.40 &1.40 &1.32 &1.24 &1.65 &1.87 &1.28 \cr
1745.2 &1.30 &1.45 &1.45 &1.27 &1.38 &1.15 &1.27 &1.50 &0.95 &1.32 \cr
1745.7 &1.22 &1.42 &1.48 &1.31 &1.32 &1.33 &1.24 &1.50 &1.94 &1.30 \cr
1746.2 &1.13 &1.35 &1.36 &1.20 &1.25 &1.35 &1.24 &1.40 &1.78 &1.35 \cr
1746.7 &1.12 &1.40 &1.52 &1.33 &1.25 &1.30 &1.30 &1.47 &0.80 &1.30 \cr
1747.2 &1.24 &1.47 &1.52 &1.40 &1.43 &1.40 &1.41 &1.55 &0.85 &1.45 \cr
1747.7 &1.55 &1.83 &1.67 &1.57 &1.60 &1.55 &1.36 &1.88 &1.70 &1.58 \cr
1748.2 &1.67 &1.80 &1.70 &1.70 &1.82 &1.63 &1.63 &1.91 &1.73 &1.67 \cr
1748.7 &1.68 &2.12 &1.83 &1.70 &1.78 &1.77 &1.50 &1.94 &1.75 &1.51 \cr
1749.2 &1.74 &1.95 &1.72 &1.67 &1.90 &1.62 &1.69 &1.88 &1.69 &1.68 \cr
1749.7 &1.33 &1.58 &1.40 &1.45 &1.43 &1.43 &1.33 &1.67 &1.47 &1.64 \cr
1750.2 &1.37 &1.67 &1.50 &1.52 &1.52 &1.25 &1.31 &1.73 &1.57 &1.46 \cr
1750.7 &1.33 &1.70 &1.77 &1.48 &1.45 &1.42 &1.17 &1.70 &1.67 &1.37 \cr
1751.2 &1.37 &1.66 &1.58 &1.48 &1.60 &1.51 &1.37 &1.83 &1.65 &1.51 \cr
1751.7 &1.72 &2.51 &1.83 &1.95 &1.98 &1.57 &0.00 &2.55 &2.05 &1.88 \cr
1752.2 &2.20 &2.42 &2.15 &2.10 &2.38 &2.03 &2.00 &2.02 &1.92 &2.07 \cr
1752.7 &1.87 &2.33 &2.25 &2.10 &2.12 &2.06 &1.90 &2.44 &1.78 &1.88 \cr
1753.2 &1.65 &1.80 &1.75 &1.80 &1.88 &1.94 &1.83 &1.98 &1.81 &2.00 \cr
1753.7 &1.60 &1.80 &1.73 &1.65 &1.70 &1.85 &1.77 &1.95 &1.00 &1.75 \cr
1754.2 &1.58 &1.62 &1.62 &1.48 &1.65 &1.83 &1.81 &2.08 &1.91 &1.80 \cr
1754.7 &1.42 &1.60 &1.58 &1.45 &1.55 &1.65 &1.59 &1.98 &1.91 &1.75 \cr
1755.2 &1.33 &1.55 &1.50 &1.42 &1.45 &1.51 &1.50 &1.72 &1.54 &1.66 \cr
1755.7 &0.00 &0.00 &0.00 &0.00 &0.00 &0.00 &0.00 &0.00 &0.00 &0.00 \cr
1756.2 &2.38 &3.05 &2.68 &2.72 &2.78 &2.63 &2.55 &2.93 &2.09 &2.39 \cr
1756.7 &1.65 &2.35 &1.99 &2.05 &2.10 &2.10 &1.67 &2.28 &2.22 &2.65 \cr
1757.2 &0.00 &0.00 &0.00 &0.00 &0.00 &0.00 &0.00 &0.00 &0.00 &0.00 \cr
1757.7 &1.43 &1.62 &1.52 &1.48 &1.48 &1.75 &1.46 &1.77 &2.22 &1.64 \cr
1758.2 &1.51 &1.84 &1.58 &1.55 &1.65 &1.97 &1.75 &2.00 &2.31 &1.79 \cr
1758.7 &1.33 &1.65 &1.62 &1.55 &1.55 &1.38 &1.32 &1.90 &2.09 &1.53 \cr
1759.2 &1.94 &1.83 &1.63 &1.52 &1.58 &1.50 &1.31 &1.80 &2.03 &1.53 \cr
1759.7 &1.44 &2.22 &1.80 &1.75 &1.62 &1.43 &1.40 &2.07 &1.71 &1.59 \cr
1760.2 &1.80 &2.22 &2.30 &2.12 &2.12 &1.83 &1.87 &2.40 &1.90 &2.06 \cr
1760.7 &1.62 &2.12 &2.15 &2.03 &1.90 &1.82 &1.98 &2.14 &1.85 &2.23 \cr
1761.2 &1.45 &1.73 &1.67 &1.45 &1.60 &1.77 &1.77 &1.73 &1.56 &1.64 \cr
1761.7 &1.43 &1.92 &1.85 &1.62 &1.62 &1.71 &1.62 &1.84 &1.59 &1.75 \cr
1762.2 &1.50 &1.85 &1.88 &1.70 &1.54 &1.60 &1.58 &1.92 &1.59 &1.74 \cr
1762.7 &1.58 &1.94 &1.92 &1.82 &1.76 &1.63 &1.63 &1.92 &1.63 &1.73 \cr
1763.2 &0.00 &0.00 &0.00 &0.00 &0.00 &0.00 &0.00 &0.00 &0.00 &0.00 \cr
1763.7 &1.60 &1.85 &1.88 &1.67 &1.72 &1.69 &1.63 &1.94 &1.77 &1.83 \cr
1764.2 &0.00 &0.00 &0.00 &0.00 &0.00 &0.00 &0.00 &0.00 &0.00 &0.00 \cr
1764.7 &0.00 &0.00 &0.00 &0.00 &0.00 &0.00 &0.00 &0.00 &0.00 &0.00 
} $$
\end{table}

\begin{table}[p]

(continued)

$$ \matrix{
\hbox{Date} & \hbox{Jiangning}& \hbox{Suzhou} & \hbox{Sungjiang} &  \hbox{Changzhou} & \hbox{Zhenjiang} & \hbox{Huaian} &
 \hbox{Yangzhou} & \hbox{Taicang} & \hbox{Haizhou} & \hbox{Tongzhou} \cr 
1765.2 &0.00 &0.00 &0.00 &0.00 &0.00 &0.00 &0.00 &0.00 &0.00 &0.00 \cr
1765.7 &1.67 &2.05 &1.93 &1.83 &1.76 &1.72 &1.67 &2.03 &2.00 &2.05 \cr
1766.2 &1.72 &2.01 &2.05 &1.85 &1.85 &1.79 &1.76 &2.15 &1.83 &1.97 \cr
1766.7 &1.65 &1.99 &2.05 &2.30 &1.77 &1.81 &1.62 &2.06 &1.99 &1.98 \cr
1767.2 &1.62 &1.70 &1.72 &1.65 &1.67 &1.79 &1.72 &1.92 &1.85 &1.82 \cr
1767.7 &0.00 &0.00 &0.00 &0.00 &0.00 &0.00 &0.00 &0.00 &0.00 &0.00 \cr
1768.2 &1.50 &1.65 &1.55 &1.45 &1.57 &1.75 &1.42 &1.70 &1.83 &1.60 \cr
1768.7 &1.75 &1.83 &1.85 &1.70 &1.72 &1.72 &1.85 &1.79 &1.84 &1.88 \cr
1769.2 &1.77 &1.82 &1.80 &1.90 &1.85 &1.95 &2.03 &1.97 &1.96 &2.10 \cr
1769.7 &1.85 &2.35 &2.08 &2.03 &2.00 &1.90 &1.83 &2.12 &1.95 &2.16 \cr
1770.2 &1.88 &1.99 &1.95 &1.83 &1.80 &1.82 &1.75 &2.15 &1.95 &1.84 \cr
1770.7 &1.55 &2.04 &1.95 &1.72 &1.76 &1.80 &1.55 &2.26 &1.96 &1.81 \cr
1771.2 &0.00 &0.00 &0.00 &0.00 &0.00 &0.00 &0.00 &0.00 &0.00 &0.00 \cr
1771.7 &1.27 &1.59 &1.77 &1.45 &1.50 &1.74 &1.41 &1.83 &1.88 &1.68 \cr
1772.2 &1.41 &1.60 &1.50 &1.45 &1.44 &1.73 &1.58 &1.67 &1.83 &1.74 \cr
1772.7 &1.30 &1.58 &1.50 &1.38 &1.35 &1.61 &1.33 &1.63 &1.84 &1.65 \cr
1773.2 &1.20 &1.39 &1.35 &1.30 &1.27 &1.45 &1.22 &1.50 &1.70 &1.38 \cr
1773.7 &1.22 &1.37 &1.42 &1.25 &1.20 &1.50 &1.15 &1.58 &1.72 &1.25 \cr
1774.2 &1.34 &1.42 &1.38 &1.30 &1.25 &1.54 &1.26 &1.58 &1.69 &1.28 \cr
1774.7 &1.55 &1.75 &1.50 &1.80 &1.67 &1.58 &1.66 &2.03 &1.71 &1.61 \cr
1775.2 &1.83 &2.07 &1.72 &1.90 &1.85 &2.13 &2.05 &2.15 &1.94 &2.15 \cr
1775.7 &2.25 &2.12 &1.90 &2.00 &2.08 &2.11 &2.40 &2.28 &1.94 &2.37 \cr
1776.2 &2.30 &2.12 &2.12 &2.30 &2.26 &2.22 &2.35 &2.40 &2.00 &2.46 \cr
1776.7 &1.65 &2.10 &2.08 &1.80 &1.80 &1.90 &1.83 &2.18 &1.96 &2.10 \cr
1777.2 &1.65 &1.95 &1.90 &1.60 &1.56 &1.88 &1.80 &2.09 &1.93 &2.08 \cr
1777.7 &1.58 &1.88 &1.95 &1.62 &1.63 &1.65 &1.53 &2.07 &1.95 &2.00 \cr
1778.2 &1.48 &1.60 &1.60 &1.38 &1.45 &1.71 &1.45 &1.84 &1.86 &1.90 \cr
1778.7 &1.90 &1.78 &1.85 &1.90 &1.78 &1.70 &1.70 &2.03 &1.92 &2.12 \cr
1779.2 &2.32 &2.57 &2.25 &2.45 &2.65 &2.35 &2.30 &2.47 &1.98 &2.35 \cr
1779.7 &2.15 &2.34 &2.22 &2.15 &2.00 &2.15 &2.00 &2.17 &2.10 &2.15 \cr
1780.2 &0.00 &0.00 &0.00 &0.00 &0.00 &0.00 &0.00 &0.00 &0.00 &0.00 \cr
1780.7 &1.50 &1.86 &1.75 &1.70 &1.65 &2.01 &1.58 &1.97 &1.97 &2.15 \cr
1781.2 &1.55 &1.60 &1.73 &1.65 &1.50 &1.96 &1.70 &1.85 &1.91 &2.15 \cr
1781.7 &1.60 &1.78 &2.03 &1.80 &1.55 &1.78 &1.78 &2.05 &1.94 &2.10 \cr
1782.2 &1.73 &2.03 &1.60 &1.99 &1.70 &2.05 &1.88 &2.08 &2.02 &2.15 \cr
1782.7 &1.60 &1.95 &1.62 &1.73 &1.58 &2.15 &1.78 &2.20 &2.08 &2.25 \cr
1783.2 &0.00 &0.00 &0.00 &0.00 &0.00 &0.00 &0.00 &0.00 &0.00 &0.00 \cr
1783.7 &0.00 &0.00 &0.00 &0.00 &0.00 &0.00 &0.00 &0.00 &0.00 &0.00 \cr
1784.2 &0.00 &0.00 &0.00 &0.00 &0.00 &0.00 &0.00 &2.10 &2.20 &2.20 \cr
1784.7 &1.85 &1.76 &1.73 &1.62 &1.65 &2.15 &2.00 &0.00 &0.00 &0.00 \cr
1785.2 &1.85 &1.77 &1.65 &1.60 &1.60 &2.24 &2.05 &2.00 &2.24 &2.25 \cr
} $$

\hrule
\vskip 0.5mm
\hrule

\vskip 2mm
Notes: The headings correspond to the names of the 10 prefectures
composing Jiangsu province. 
The decimals .2 and .7 in the date respectively refer to 
the second and eighth month of the Chinese lunar
calendar; a lunar month
being shorter than 30 or 31 days some years had 13 months.
In the western calendar those dates would approximately correspond to
February and August. 
The prices are expressed in taels per shi 
(a shi is 103.5 liters).
0.00 means ``missing figure'';
between 1786 and 1792 there are many missing data
which is why these years were omitted. 
Source: see text.

\end{table}

\normalsize


 \scriptsize

\begin{table}[p]

\centerline{\bf Table A2 \ Wheat prices in France: 1756-1790}
\vskip 3mm
\hrule
\vskip 0.5mm
\hrule
\vskip 2mm
$$ \matrix{
\tvi \hbox{Year} & \hbox{Alen\c con} & \hbox{Amiens} & 
\hbox{Bourges} & \hbox{Bourgogne} & \hbox{Bretagne} &
\hbox{Caen} & \hbox{Lyon} & \hbox{Riom} & \hbox{Rouen} & \hbox{Tours} \cr 
1756 &1615 &1380 &1405 &1720 &1704 &1670 &1785 &1455 &1575 &1410 \cr
1757 &2400 &2520 &1330 &2010 &2020 &2385 &1804 &1670 &2905 &1565 \cr
1758 &1835 &1485 &1505 &2230 &1715 &1854 &1985 &1760 &1850 &1505 \cr
1759 &1750 &1385 &1590 &2165 &1800 &1585 &2210 &1870 &1890 &1490 \cr
1760 &1685 &1520 &1410 &2145 &1950 &1555 &2075 &1670 &1950 &1365 \cr
1761 &1555 &1335 &2075 &1650 &1935 &1570 &1690 &1390 &1580 &1305 \cr
1762 &1860 &1615 &1215 &1570 &1800 &1854 &1610 &1310 &1745 &1445 \cr
1763 &1415 &1335 &1005 &1485 &1525 &1505 &1335 &1355 &1430 &1425 \cr
1764 &1235 &1320 &1170 &1679 &1625 &1115 &1640 &1420 &1405 &1305 \cr
1765 &1595 &1604 &1335 &1675 &1650 &1710 &1979 &1625 &1675 &1625 \cr
1766 &1785 &1635 &1935 &2485 &2245 &1925 &2495 &2210 &2035 &2130 \cr
1767 &2190 &2285 &1940 &2775 &2180 &2015 &2655 &2590 &2400 &1715 \cr
1768 &2780 &2920 &1920 &2505 &2545 &2510 &2230 &2065 &3015 &2190 \cr
1769 &2960 &2270 &2035 &2680 &2480 &2360 &2355 &2200 &2730 &2665 \cr
1770 &3440 &2355 &3405 &3505 &3130 &3230 &3205 &3480 &2990 &3015 \cr
%
1771 &2865 &2620 &2795 &3770 &2630 &2730 &3655 &3559 &2850 &2305 \cr
1772 &2610 &3375 &2695 &2655 &2825 &2710 &3020 &3120 &2650 &2770 \cr
1773 &2695 &2460 &1950 &2625 &2710 &2495 &2885 &2485 &2850 &2355 \cr
1774 &2280 &2095 &1715 &2330 &2335 &1950 &2625 &2255 &2320 &1954 \cr
1775 &2785 &2425 &2350 &2590 &3145 &2555 &2630 &2560 &2765 &2355 \cr
1776 &2345 &1845 &1750 &1904 &2310 &2000 &2035 &1875 &2515 &2105 \cr
1777 &2475 &1950 &1679 &1765 &2090 &2165 &2020 &2135 &2520 &1979 \cr
1778 &2140 &1790 &1665 &2265 &2305 &2075 &2630 &2670 &2215 &1895 \cr
1779 &2225 &1590 &1690 &2595 &2055 &2405 &2875 &2410 &2095 &1650 \cr
1780 &2180 &1515 &1640 &2285 &2095 &2395 &2375 &2000 &2080 &1610 \cr
1781 &2220 &1750 &1790 &2120 &2240 &2195 &2180 &1925 &2230 &1910 \cr
1782 &2180 &1645 &2475 &2470 &2850 &2385 &2555 &2415 &1960 &2420 \cr
1783 &2135 &1725 &2340 &2745 &2555 &2205 &2965 &2540 &1975 &2305 \cr
1784 &2720 &2345 &2070 &2470 &2570 &2540 &2525 &2145 &2875 &2320 \cr
1785 &2455 &1860 &1925 &2285 &2745 &2605 &2235 &1865 &2195 &2485 \cr
1786 &2305 &1660 &1915 &1954 &2915 &2485 &2190 &1760 &1920 &2620 \cr
1787 &2260 &1735 &2100 &2210 &2190 &2060 &2450 &1960 &2135 &1979 \cr
1788 &2390 &2100 &2310 &2710 &2240 &2290 &2735 &2395 &2495 &2405 \cr
1789 &3240 &3375 &3440 &3370 &3110 &3215 &3634 &3340 &3384 &3100 \cr
1790 &2790 &2140 &3270 &3215 &3060 &2750 &3365 &3209 &2750 &2925 
} $$

\hrule
\vskip 0.5mm
\hrule

\vskip 2mm
Notes: The prices are expressed in hundredth of livres per ``setier de 
Paris'' (a unit of volume equal to 156 liters and equivalent to a 
weight of about 120 kilogram of wheat). The city names 
refer to the ``G\'en\'eralit\'es'' (i.e. districts)
of which the corresponding cities were the centers. 
Source: Labrousse 1933.

\end{table}

\normalsize

\count101=0

\ifnum\count101=1

 \scriptsize

\begin{table}[htb]

\centerline{\bf Table A3 \ Wheat prices in France: 1806-1900}
\vskip 3mm
\hrule
\vskip 0.5mm
\hrule
\vskip 2mm
$$ \matrix{
\tvi \hbox{Year} & \hbox{Alen\c con} & \hbox{Amiens} & 
\hbox{Bourges} & \hbox{Bourgogne} & \hbox{Bretagne} &
\hbox{Caen} & \hbox{Lyon} & \hbox{Riom} & \hbox{Rouen} & \hbox{Tours} \cr 
1806 &1575 &1606 &1552 &1731 &1075 &1560 &2109 &1648 &1555 &1364 \cr
1807 &1750 &1680 &1603 &1735 &1226 &1692 &2040 &1480 &1682 &1596 \cr
1808 &1633 &1459 &1403 &1486 &1089 &1573 &1772 &1428 &1495 &1370 \cr
1809 &1474 &1193 &1283 &1413 &1112 &1589 &1571 &1568 &1396 &1027 \cr
1810 &1866 &1614 &1597 &1915 &1759 &2176 &2172 &2140 &1827 &1253 \cr
1811 &2126 &2011 &2317 &2659 &1725 &2312 &2974 &2452 &2242 &1954 \cr
1812 &3554 &3299 &3572 &3228 &3056 &3558 &3619 &3152 &3700 &3238 \cr
1813 &2231 &2139 &2254 &2107 &1895 &2431 &2325 &1962 &2465 &1831 \cr
1814 &1499 &1366 &1576 &1560 &1372 &1577 &1899 &1503 &1531 &1389 \cr
1815 &1612 &1488 &1634 &1715 &1595 &1612 &2016 &1780 &1636 &1556 \cr
1816 &2564 &2714 &2253 &2795 &2320 &2685 &2994 &2619 &2923 &2111 \cr
1817 &3257 &3687 &3075 &4005 &2778 &3426 &4099 &3478 &3697 &2910 \cr
1818 &2315 &2091 &2273 &2198 &2265 &2311 &2582 &2102 &2269 &2265 \cr
1819 &1862 &1620 &1656 &1571 &1945 &1973 &1890 &1667 &1860 &1632 \cr
1820 &2155 &1850 &1619 &1725 &1840 &2416 &2008 &1824 &2204 &1709 \cr
1821 &2023 &1780 &1558 &1585 &1749 &2162 &1834 &1633 &2109 &1608 \cr
1822 &1520 &1418 &1179 &1362 &1399 &1660 &1599 &1403 &1587 &1259 \cr
1823 &1696 &1656 &1337 &1609 &1473 &1822 &1875 &1564 &1734 &1519 \cr
1824 &1549 &1348 &1358 &1556 &1468 &1719 &1794 &1495 &1615 &1424 \cr
1825 &1686 &1531 &1274 &1619 &1548 &1877 &1828 &1442 &1681 &1285 \cr
1826 &1848 &1528 &1328 &1577 &1546 &1906 &1722 &1412 &1727 &1432 \cr
1827 &1756 &1691 &1737 &2009 &1513 &1802 &2197 &1875 &1891 &1568 \cr
1828 &2164 &2207 &1988 &2277 &1828 &2198 &2598 &2325 &2362 &1883 \cr
1829 &2646 &2571 &2118 &2044 &2144 &2675 &2275 &2022 &2550 &2305 \cr
1830 &2217 &2071 &1978 &2473 &1866 &2308 &2734 &2448 &2098 &2017 \cr
1831 &2090 &2191 &1943 &2128 &2001 &2209 &2433 &2237 &2211 &1999 \cr
1832 &2120 &2110 &1689 &2141 &1942 &2178 &2332 &2154 &2092 &1902 \cr
1833 &1554 &1466 &1371 &1742 &1370 &1550 &1895 &1636 &1517 &1381 \cr
1834 &1502 &1392 &1292 &1417 &1404 &1495 &1603 &1436 &1485 &1364 \cr
1835 &1463 &1395 &1264 &1402 &1446 &1458 &1486 &1393 &1445 &1390 \cr
1836 &1516 &1389 &1510 &1682 &1639 &1690 &1800 &1749 &1520 &1479 \cr
1837 &1785 &1486 &1765 &1785 &1595 &1932 &1773 &1679 &1627 &1739 \cr
1838 &2095 &2011 &1880 &1959 &1716 &2128 &1964 &1719 &2029 &1904 \cr
1839 &2201 &2260 &2181 &2345 &1958 &2245 &2476 &2099 &2278 &2146 \cr
1840 &2387 &2254 &2119 &2286 &2055 &2528 &2450 &2036 &2362 &2108 \cr
1841 &1711 &1656 &1618 &1945 &1609 &1839 &2148 &1892 &1834 &1624 \cr
1842 &1784 &1866 &1714 &2004 &1643 &1952 &2172 &1833 &1975 &1692 \cr
1843 &2027 &1834 &1819 &2048 &1716 &2102 &2183 &1999 &2022 &1876 \cr
1844 &1942 &1766 &1852 &1823 &1717 &2043 &1981 &2038 &1959 &1855 \cr
1845 &1800 &1684 &1855 &1843 &1704 &1888 &1957 &1950 &1903 &1801 \cr
1846 &2219 &2236 &2310 &2501 &1953 &2271 &2597 &2478 &2457 &2222 \cr
1847 &3143 &3001 &3147 &2873 &2562 &3174 &2857 &3247 &3021 &3123 \cr
1848 &1608 &1451 &1425 &1611 &1556 &1705 &1805 &1628 &1582 &1480 \cr
1849 &1572 &1498 &1208 &1459 &1486 &1739 &1573 &1396 &1648 &1407 \cr
1850 &1412 &1378 &1119 &1333 &1358 &1458 &1502 &1261 &1450 &1316 \cr
1851 &1314 &1439 &1134 &1363 &1337 &1362 &1500 &1218 &1436 &1302 \cr
1852 &1654 &1729 &1428 &1767 &1605 &1743 &1816 &1531 &1755 &1529 \cr
1853 &2339 &2318 &2024 &2291 &2056 &2413 &2325 &1967 &2295 &2194 \cr
1854 &3043 &2828 &2732 &2721 &2846 &3154 &2888 &2690 &2924 &2770 \cr
1855 &3156 &3130 &2775 &2874 &2804 &3330 &3068 &2921 &3242 &2839 \cr
1856 &2989 &2920 &3052 &2918 &2665 &3070 &3218 &3294 &3085 &2931 \cr
1857 &2271 &2043 &2333 &2267 &2188 &2287 &2470 &2523 &2260 &2317 \cr
1858 &1643 &1488 &1453 &1578 &1543 &1679 &1665 &1635 &1592 &1477 \cr
1859 &1614 &1573 &1379 &1606 &1614 &1702 &1677 &1561 &1663 &1489 \cr
1860 &2259 &2091 &1791 &1979 &2004 &2345 &2019 &1849 &2188 &1936 \cr
1861 &2684 &2381 &2312 &2346 &2299 &2734 &2337 &2332 &2534 &2565 \cr
1862 &2346 &2155 &2114 &2123 &2248 &2479 &2234 &2455 &2284 &2178 \cr
1863 &1855 &1835 &1771 &1892 &1789 &1933 &1956 &1895 &1885 &1818 \cr
1864 &1780 &1622 &1584 &1681 &1625 &1759 &1718 &1640 &1706 &1613 \cr
1865 &1674 &1577 &1447 &1549 &1516 &1693 &1638 &1460 &1688 &1494 \cr
1866 &2009 &1887 &1720 &1902 &1857 &2059 &1944 &1786 &2089 &1882 \cr
1867 &2742 &2577 &2519 &2437 &2548 &2858 &2614 &2547 &2736 &2634 \cr
1868 &2706 &2432 &2566 &2441 &2475 &2682 &2497 &2592 &2530 &2704 \cr
1869 &2020 &1826 &1908 &1946 &1850 &2010 &1982 &1953 &1964 &1931 \cr
1870 &2174 &1894 &1950 &1990 &1897 &2137 &2006 &1935 &2085 &1997 \cr
} $$

\hrule
\vskip 0.5mm
\hrule

\vskip 2mm

\end{table}

\vfill \eject

\begin{table}[htb]

\centerline{\bf Table A3 \ Wheat prices in France: 1806-1900 (continued)}
\vskip 3mm
\hrule
\vskip 0.5mm
\hrule
\vskip 2mm
$$ \matrix{
\tvi \hbox{Year} & \hbox{Alen\c con} & \hbox{Amiens} & 
\hbox{Bourges} & \hbox{Bourgogne} & \hbox{Bretagne} &
\hbox{Caen} & \hbox{Lyon} & \hbox{Riom} & \hbox{Rouen} & \hbox{Tours} \cr
1871 &2847 &2673 &2678 &2587 &2262 &2655 &2443 &2580 &2674 &2934 \cr
1872 &2462 &2435 &2119 &2401 &2264 &2698 &2235 &2355 &2310 &2277 \cr
1873 &2749 &2679 &2508 &2759 &2492 &2818 &2573 &2744 &2633 &2554 \cr
1874 &2525 &2493 &2462 &2521 &2443 &2702 &2322 &2693 &2397 &2543 \cr
1875 &1899 &1910 &1780 &1975 &1743 &1872 &1844 &1997 &1888 &1836 \cr
1876 &2088 &2093 &1953 &2195 &1954 &2168 &1945 &2051 &2081 &1976 \cr
1877 &2507 &2469 &2267 &2363 &2443 &2634 &2215 &2306 &2428 &2335 \cr
1878 &2321 &2204 &2260 &2421 &2152 &2375 &2172 &2441 &2174 &2375 \cr
1879 &2173 &2096 &2107 &2445 &2051 &2260 &1971 &2378 &2019 &2216 \cr
1880 &2304 &2135 &2279 &2484 &2200 &2360 &2368 &2554 &2085 &2355 \cr
1881 &2281 &2224 &2084 &2356 &1754 &2308 &2150 &2368 &2110 &2266 \cr
1882 &2205 &2103 &2035 &2223 &2015 &2257 &2161 &2313 &2136 &2213 \cr
1883 &1939 &1790 &1715 &1940 &1779 &1936 &1884 &1959 &1838 &1897 \cr
1884 &1826 &1709 &1594 &1829 &1559 &1862 &1759 &1879 &1719 &1752 \cr
1885 &1688 &1643 &1572 &1739 &1525 &1766 &1655 &1718 &1644 &1633 \cr
1886 &1692 &1617 &1580 &1740 &1576 &1685 &1712 &1732 &1667 &1671 \cr
1887 &1863 &1693 &1716 &1838 &1497 &1836 &1825 &1898 &1741 &1817 \cr
1888 &2029 &1872 &1825 &1824 &1817 &2019 &1817 &1914 &1893 &1907 \cr
1889 &1842 &1681 &1761 &1861 &1720 &1945 &1815 &2050 &1717 &1826 \cr
1890 &1967 &1825 &1869 &1851 &1847 &2021 &1854 &2000 &1870 &1936 \cr
1891 &2153 &1989 &2083 &2098 &1909 &2150 &2074 &2246 &2031 &2218 \cr
1892 &1750 &1690 &1782 &1851 &1549 &1737 &1796 &2052 &1712 &1976 \cr
1893 &1576 &1553 &1593 &1675 &1335 &1599 &1646 &1720 &1533 &1728 \cr
1894 &1502 &1436 &1455 &1510 &1522 &1503 &1500 &1537 &1455 &1531 \cr
1895 &1400 &1391 &1354 &1430 &1420 &1439 &1422 &1385 &1405 &1333 \cr
1896 &1439 &1436 &1422 &1527 &1258 &1463 &1493 &1551 &1399 &1466 \cr
1897 &1939 &1860 &1936 &1953 &1793 &1914 &1888 &1894 &1818 &2005 \cr
1898 &1975 &1942 &2047 &2090 &1906 &2026 &2016 &2017 &1856 &2093 \cr
1899 &1494 &1470 &1489 &1533 &1372 &1487 &1519 &1560 &1473 &1521 \cr
1900 &1504 &1445 &1456 &1477 &1406 &1480 &1443 &1524 &1455 &1468 
} $$

\hrule
\vskip 0.5mm
\hrule

\vskip 2mm
Notes: The prices are expressed in centimes per hectoliters
(remember that a centime is one hundredth of a franc
and that throughout the 19th century one franc
was almost equal (in gold content) to one livre. The location names
should be interpreted as the centers of the corresponding regions; the
actual limits of the administrative divisions
were redefined during the Revolution and are thus not exactly 
identical to those referred to in table A2.
Source: 1806-1870: Labrousse (1970); 1871-1900: Drame et al (1991).

\end{table}

\normalsize

\fi

\null

\vfill \eject

\centerline{\bf \Large References}

\vskip 1cm

\qparr
Archives Statistiques du Minist\`ere des Travaux Publics, 
de l'Agriculture et du Commerce 1837. [Statistical archives of the 
Department of Agriculture and Commerce]
Paris.

\qparr
Brenner (Y.S.), Kaelble (H.), Thomas (M.) eds. 1991: Income
distribution in historical perspective. Cambridge University 
Press. Cambridge. 

\qparr
Chuan (H.-S.), Kraus (R.A.) 1975: Mid-Ch'ing rice markets and trade: 
an essay in price history. Harvard University Press. Cambridge (Ma).

\qparr
Cuddington (J.T.), Urzua (C.M.) 1989: Trends and cycles in the net
barter terms of trade: a new approach. 
The Economic Journal, June, 426-442.

\qparr
Dewald (W.G.), Thursby (J.G.), Anderson (R.G.) 1986: Replication
in empirical economics. American Economic Review 76,587-603.

\qparr
Doyle (A.C.) 1975: The adventure of the crooked man. in: The complete
adventures and memoirs of Sherlock Holmes: a facsimile of the
original Strand Magazine stories 1891-1893. C.N. Potter. New York. 

\qparr
Drame (S.), Gonfalone (C.), Miller (J.A.), Roehner (B.) 1991:
Un si\`ecle de commerce du bl\'e en France 1825-1913. 
[Wheat trade and wheat prices in France 1825-1913]. Economica. Paris.

\qparr
Hagen (J. von) 1989: Relative commodity prices and cointegration.
Journal of Business and Economic Statistics 7,4,497-503.

\qparr
Kendall (M.), Stuart (A.), Ord (J.K.) 1987: Advanced theory of 
statistics. Vol.1. Charles Griffin. London.

\qparr
Jacobs (A.), Richter (H.) 1935: Die Gro\ss handelspreise in
Deutschland von 1792 bis 1934 [Wholesale prices in Germany
from 1792 to 1934]. Hanseatische Verlagsanstalt. Berlin.

\qparr
Labrousse (C.-E.) 1933: Esquisse du mouvement des prix et des
revenus en France au XVIIIe si\`ecle. [The course of prices and
income in France in the 18th century]
Dalloz. Paris.

\qparr
Labrousse (E.), Romano (R.), Dreyfus (F.-G.) 1970: Le prix
du froment en France au temps de la monnaie stable (1726-1913).
[Wheat prices in France: (1726-1913)]. SEVPEN. Paris.

\qparr
Mj\"oset (L.) et al. eds. 1997: Methodological issues in 
comparative social science. Comparative Social Research Vol.16. Jai 
Press. Greenwich (Conn.)

\qparr
OECD 1960, 1989: Main economic indicators. Historical statistics. 

\qparr
Prebisch (R.) 1950: The economic problem of Latin America and its
principal problems. United Nations. Lake Success. 

\qparr
Roehner (B.M.) 1995: Theory of markets. Springer-Verlag. Berlin. 

\qparr
Roehner (B.M.) 1996: The role of transportation costs in the 
economics of commodity markets. American Journal of Agriculture
Economics 78,339-353.

\qparr
Roehner (B.M.) 2000: The correlation length of commodity markets.
Theoretical framework. European Physical Journal B 13,189-200.

\qparr
Singer (H.) 1950: The distributions of gains between investing and 
borrowing countries. American Economic Review. Papers and 
Proceedings, 40, 473-485.

\qparr
Smeeding (T.M.), O'Higgins (M.), Rainwater (L.) 1990: Poverty,
inequality and income in comparative perspective. The Luxembourg
Income Study (LIS). The Urban Institute Press. Washington. 

\qparr
Wilkinson (E.) 1969: The nature of Chinese grain price quotations,
1600-1900. Transactions of the International Conference of Orientalists
14, 54-65. 

\qparr
Zellner (A.) 1988: Causality and causality laws in economics.
Journal of Econometrics 39,7-21.

\vfill \eject

{\bf Captions of the figures}
\qpar

Fig.1 (a) Operational definition of the correlation length. The rectangles
schematize grain markets; the interactions between them is represented by 
the dots; a the level of 
price series they give rise to correlations $ c_{12}, c_{13}, 
c_{14}, \ldots $. (b) Once these correlations are plotted 
(on a logarithmic scale) as a function of
distance the correlation length $ \delta $ can be read from the slope $ m $ 
of the regression line as: $ \delta =-1/m $. 
\qpar

Fig.2a Map of South-East China. Within South-East China the provinces of
Jiangsu and Yangzi Delta are the main plain regions. 
\qpar

Fig.2b Map of France with location of markets.
\qpar

Fig.3 Correlation as a function of distance in France 1756-1790. 
The 
corresponding graph for China would be very similar in shape and has
therefore been omitted. 
\qpar

Fig.4a Trend for market integration in Yangzi Delta 1742-1794. The 
vertical scale corresponds to the Max-Min variable referred to as
$ a $ in the text.
\qpar

Fig.4b Trend for market integration in France 1756-1900. There is 
an acceleration in the improvement of market integration between
the 18th and 19th century. The vertical scale corresponds to the
Max-Min variable defined in the text. 

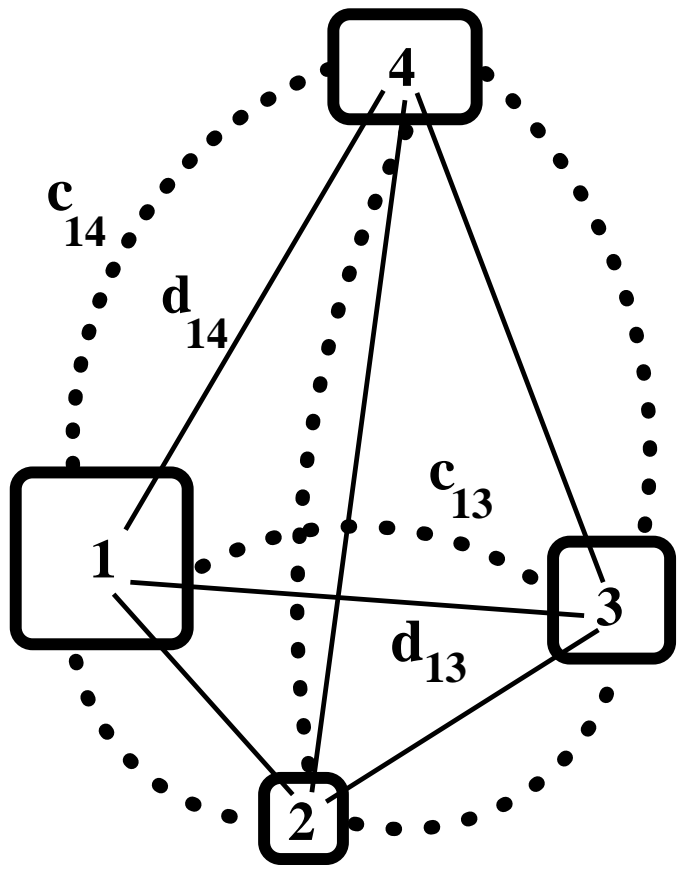
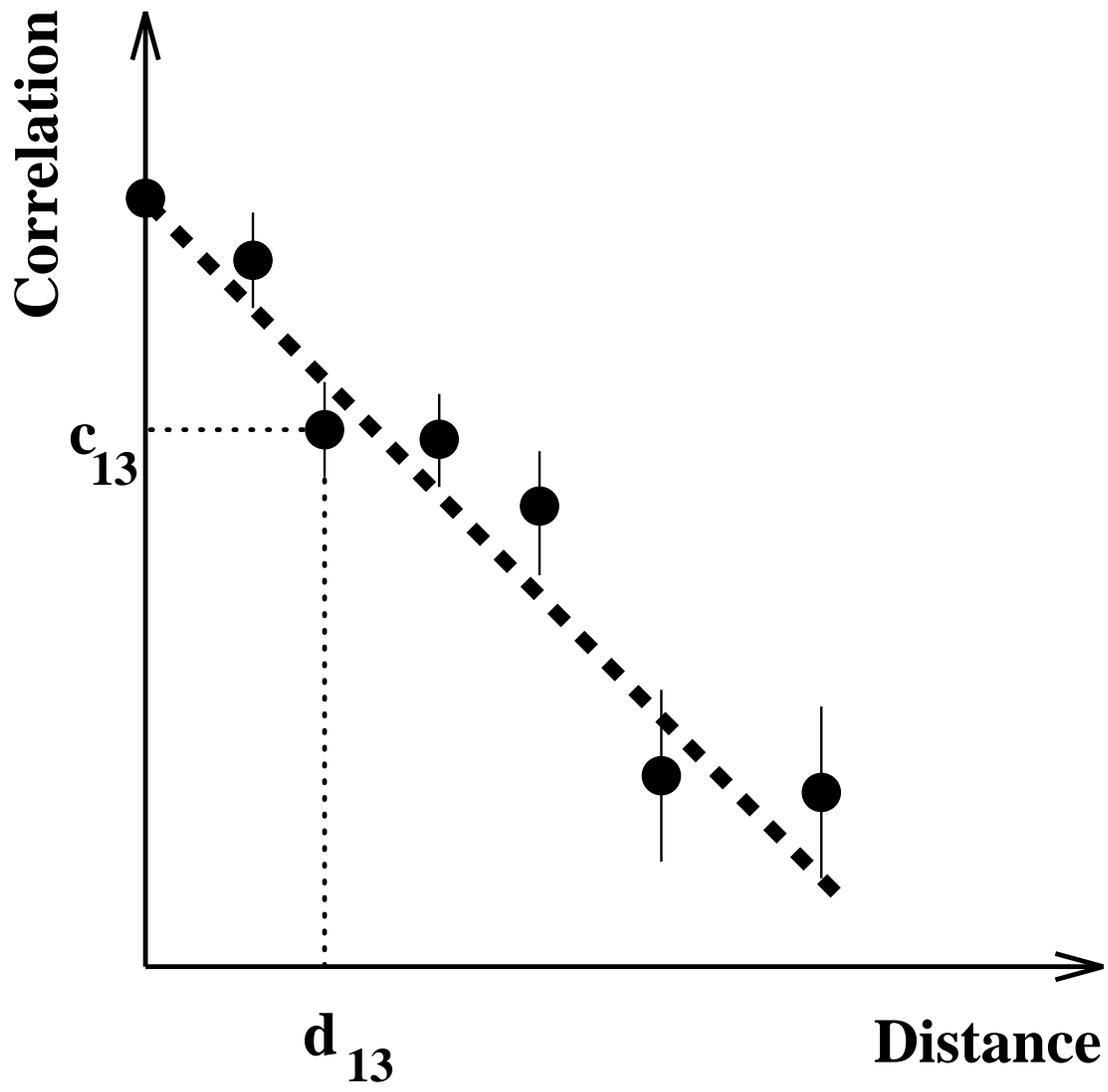
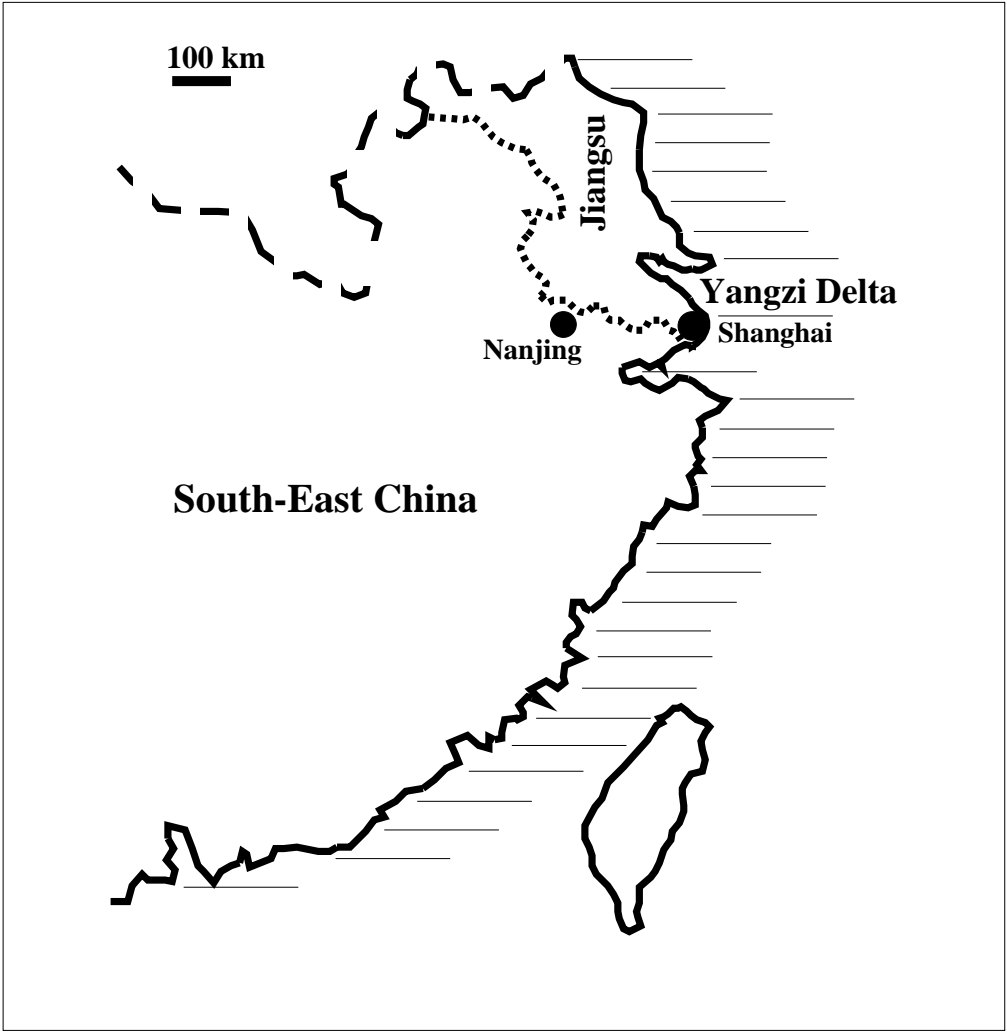
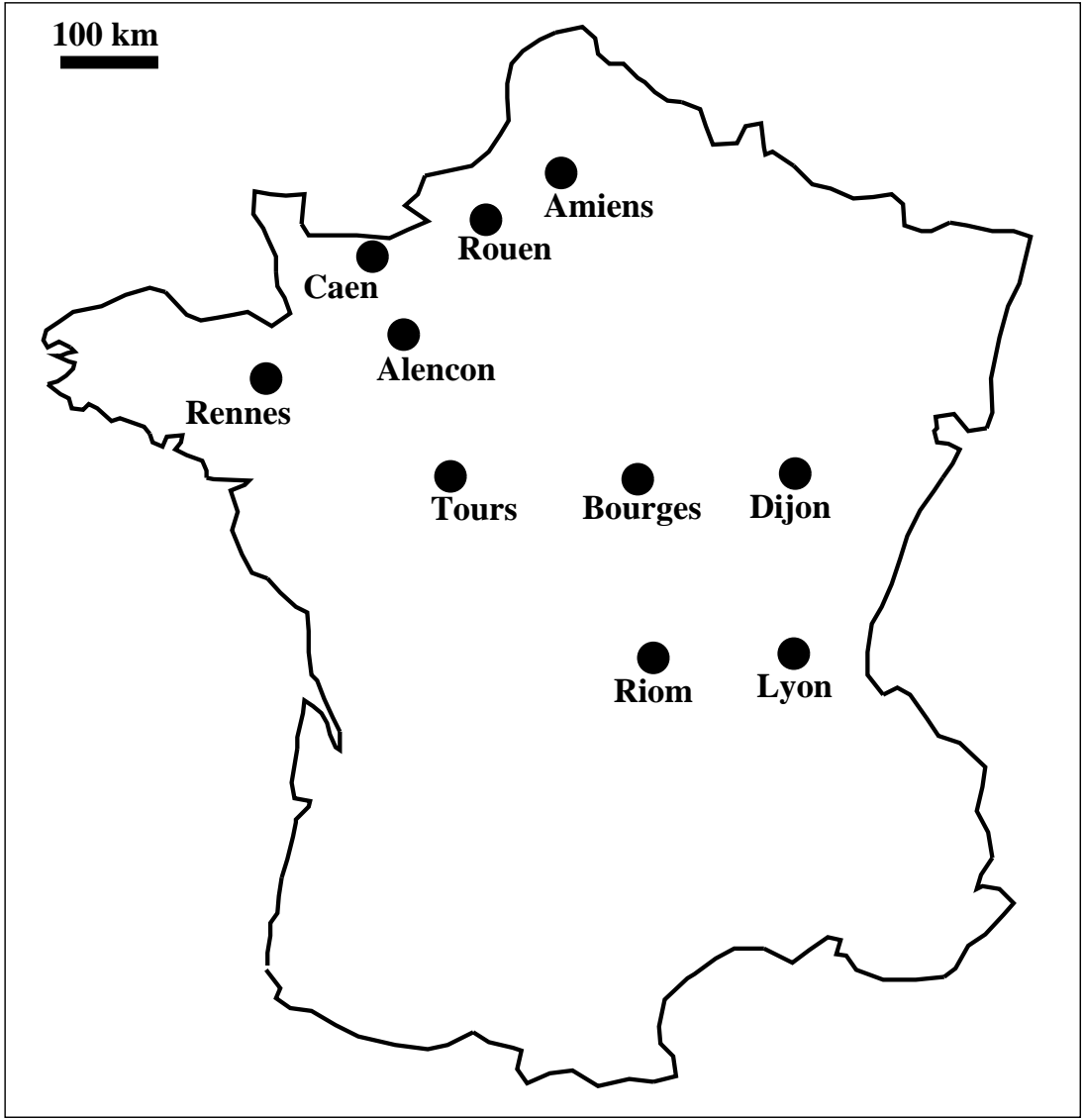
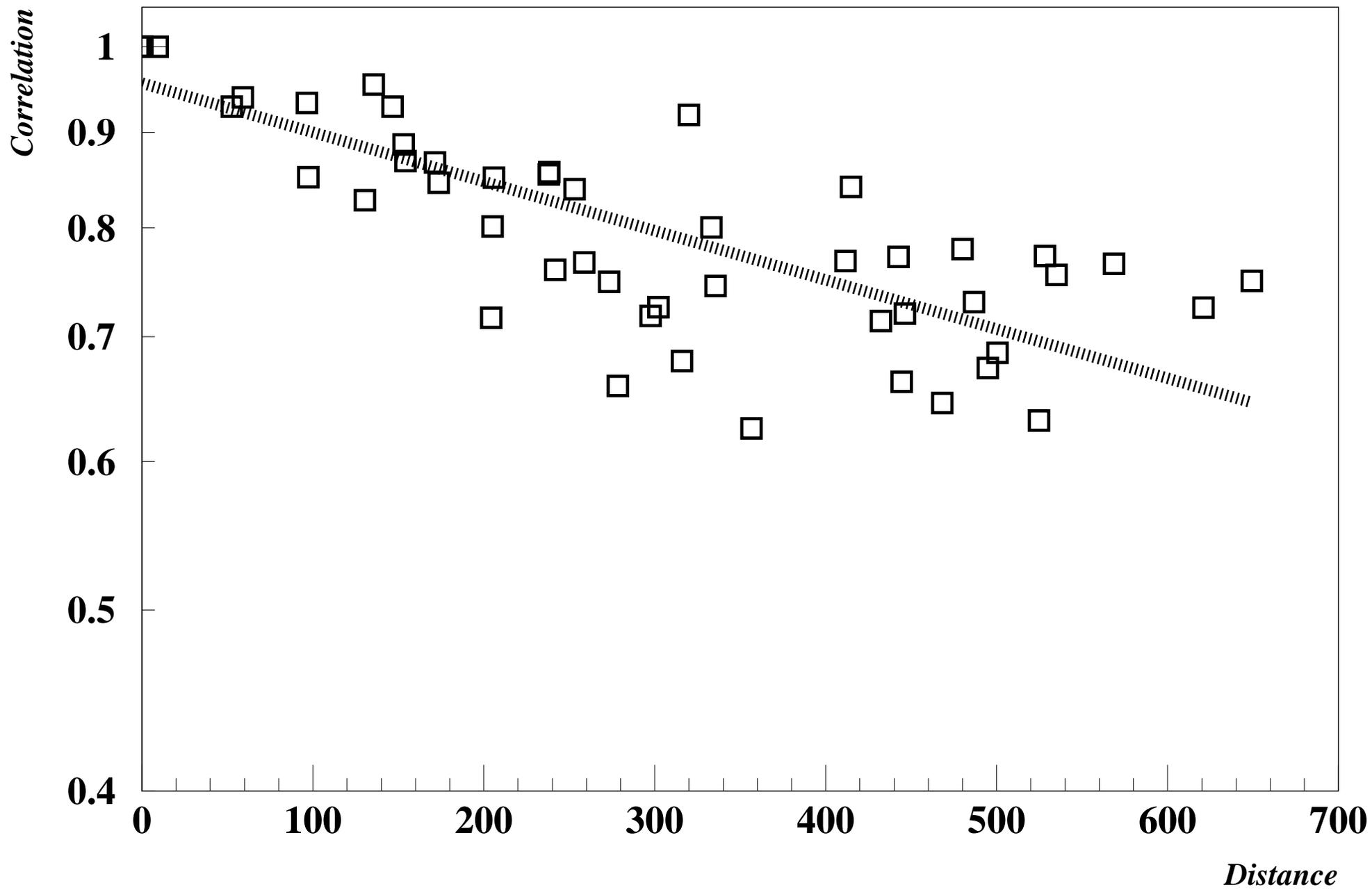
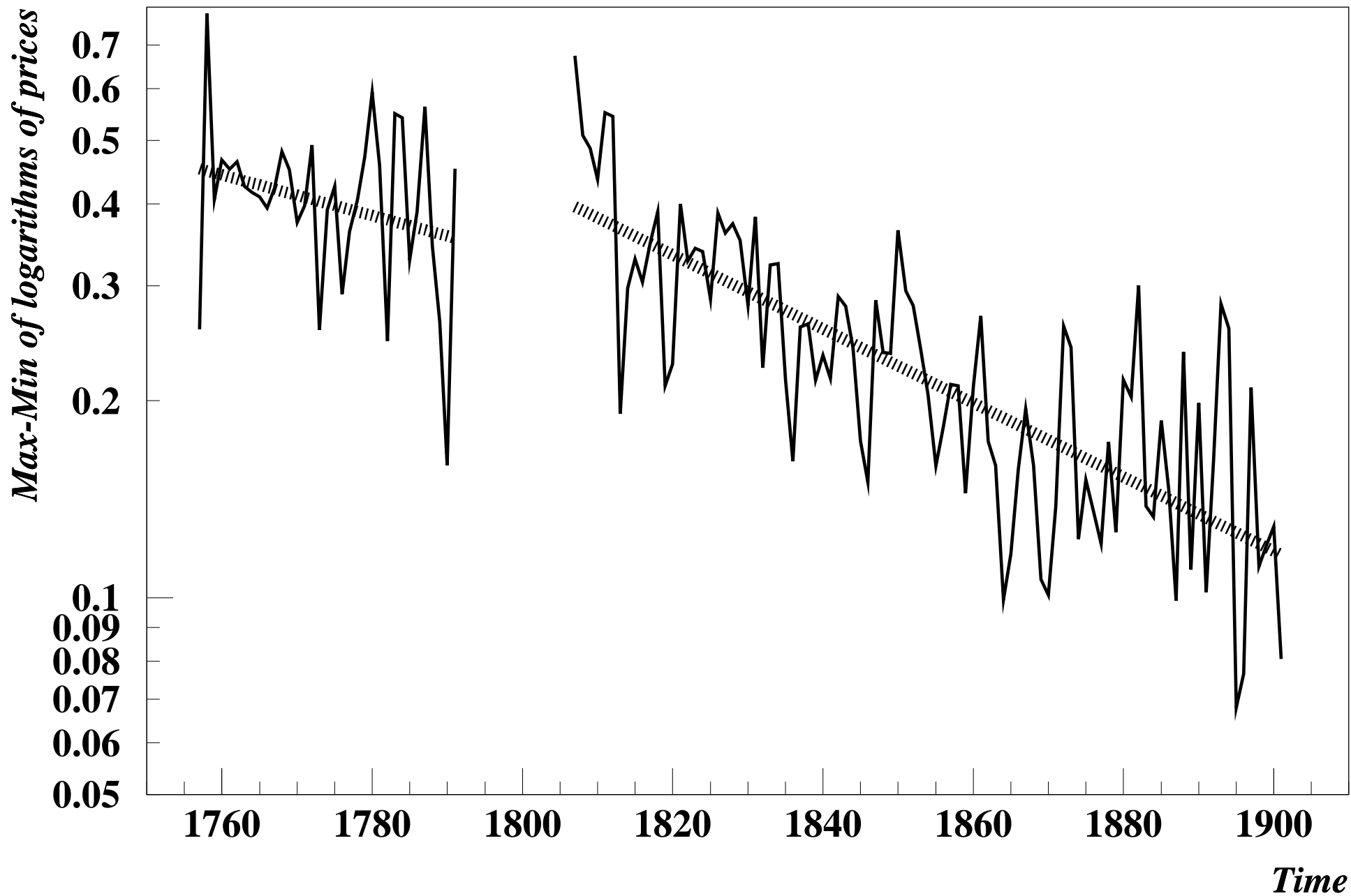
\end{document}